\documentclass[12pt]{article}
\usepackage{pic04}
\usepackage{hyperref}
\usepackage{url}
\ifx\pdftexversion\undefined
  \usepackage[dvips]{graphicx}
\else
  \usepackage[pdftex]{graphicx}
\fi

\newcommand{\beq}{\begin{equation}}
\newcommand{\eeq}{\end{equation}}
\newcommand{\bea}{\begin{eqnarray}}
\newcommand{\eea}{\end{eqnarray}}
\newcommand{\bi}{\bibitem}

\newcommand{\Dslash}{\rlap{/}\kern-2.0pt D}

\begin{document}

\title{\bf LATTICE QCD AND FLAVOR PHYSICS}
\author{
Matthew Wingate        \\
{\em Institute for Nuclear Theory, University of Washington, Seattle, WA
98195-1550}}
\maketitle

%
%
\begin{figure}[h]
\begin{center}
%
%
%
%
\vspace{4.5cm}
\end{center}
\end{figure}

\baselineskip=14.5pt
\begin{abstract}
Now that lattice QCD simulations are able to include effects of
light sea quarks, the prospects are good for constraining quark
flavor phenomenology.
This review talk for particle physics experimentalists begins with
an introduction intended to describe broadly the steps of lattice
Monte Carlo simulations.  The remainder of the talk is a brief
survey of recent and ongoing calculations relevant for quark
flavor physics.
\end{abstract}
\newpage

\baselineskip=17pt

\section{Introduction}
\label{sec:intro}

In principle lattice QCD is an {\it ab initio} method for numerically
computing the QCD spectrum and many hadronic matrix elements.
In practice there are several difficulties which must be overcome.  
Getting to the point where
all uncertainties in the calculations can be reduced systematically 
has taken (at least) 2 decades.  Are we there yet?  Possibly.
Recently it has been shown that the inclusion of light sea quark effects,
via an improved staggered fermion action,
removes the uncontrollable errors of the quenched approximation. 
This allows for more accurate investigation of quark mass extrapolations
as well as finite volume and discretization effects.

Given that the dominant uncertainty for most of the constraints
on the CKM parameters $\rho$ and $\eta$ comes from hadronic transitions
\cite{Charles:2004jd,Bona:2004sj}, little more needs to be said to
motivate lattice QCD calculations.  Several speakers at this conference
have already remarked on the importance of lattice results for 
flavor physics phenomenology, and Lubicz talked at length about the 
impact that lattice QCD results have in fits to the CKM parameters 
at the Lattice Field Theory Symposium this year\cite{Lubicz:Lat2004}.  
Furthermore, the CLEO-c experiment will make measurements which will 
allow for greater tests of lattice phenomenology 
\cite{Shipsey:Lat2004,Galik:PiC2004}.

Since this is a conference mainly for experimentalists, I will give an
introduction to lattice QCD calculations which touches upon
important ingredients but skips many details not 
related to the rest of the talk.  The quenched approximation and the 
use of improved staggered quarks to unquench are of particular relevance.
The second half of the talk presents a selection of results,
some preliminary, which are important for high energy phenomenology.
Due to time and space constraints, this presentation is focused by
the lens of my interests.
A more thorough review of recent results is 
in preparation \cite{Wingate:Lat2004}.

\section{Lattice Monte Carlo Calculations}
\label{sec:monte}

The goal of this section is to give the nonexpert a broad outline
of lattice QCD simulations and to highlight where recent
progress has made a substantial improvement.  The interested reader
may find more details in reviews such as Ref.~\cite{DeGrand:2003xu}
(and the many references therein).

QCD is a strongly
coupled theory at energy scales below approximately 1~GeV: processes
with 23 gluon exchanges are just as important as a single gluon exchange.
Matrix elements involving low energy hadrons cannot be calculated
directly using a perturbative expansion about small coupling.  
Instead a nonperturbative way of evaluating path integrals is needed.
Lattice field theory provides this, and moreover it
is the only known way to regulate a quantum field
theory nonperturbatively.  By rotating the time coordinate $90^\circ$
in the complex plane (``going to Euclidean spacetime'') and working 
with a discrete and finite spacetime, the path integral
representations of hadronic matrix elements can be computed numerically
by Monte Carlo simulation.

\subsection{Generating Important Samples of Glue}

\begin{figure}[t]
\begin{center}
\includegraphics[width=12cm]{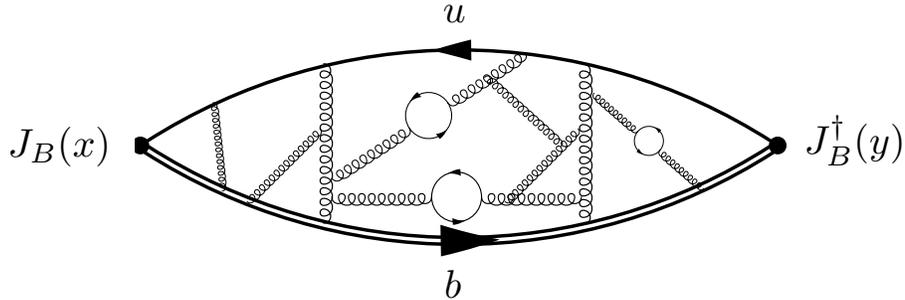}
\caption{\label{fig:corr} {\it Cartoon of a
$B^-$ meson propagator.  The $b$ and $\overline{u}$ quarks
propagate through nonperturbative glue.}}
\end{center}
\end{figure}

Monte Carlo simulations of lattice QCD rely on the same principles
used for simulations of classical statistical mechanical models, 
{\it e.g.}\ the Ising model.  Recall that any quantum 
observable, like the propagator of some hadron from $x$ to $y$
({\it e.g.}\ Fig.~\ref{fig:corr}), 
can be computed from the path integral, 
\beq
\langle J^\dagger(y)J(x) \rangle 
~=~ \frac{1}{\cal Z}\int [d\psi][d\overline{\psi}]
[d U] ~J^\dagger(y)J(x)~ e^{i {\cal S}_M}
\eeq
where $\psi$ represents the quark field and $U$ the glue field, 
and  ${\cal Z} \equiv \langle 1 \rangle$.
$J$ creates ($J^\dagger$ annihilates) the hadron, and all possible
paths are included with a phase determined
by the action ${\cal S}_M = \int d^3 x' \, dt' ~{\cal L}_M$, 
where ${\cal L}_M$ is the QCD Lagrangian in Minkowski spacetime.
If we change from physical Minkowski spacetime to Euclidean spacetime by
introducing an imaginary time coordinate $\tau \equiv it$, then
\beq
\langle J^\dagger(y)J(x) \rangle ~=~ \frac{1}{Z} 
\int [d\psi][d\overline{\psi}]
[d U] ~J^\dagger(y)J(x)~ e^{-\int d^3 x \,d\tau {\cal L}_E} \,.
\label{eq:EuclidEV}
\eeq
Let us compare (\ref{eq:EuclidEV}) to the expectation value for the 
magnetization in the Ising model
\beq
\langle s_i \rangle 
~=~ \frac{1}{Z'}  \sum_{s_i = \pm 1} s_i
~e^{-H/T} \, .
\label{eq:IsingEV}
\eeq
$Z'$ is the corresponding partition function.
One can identify the Ising spins with the fermion and gluon 
degrees of freedom, and the statistical mechanical Hamiltonian
with the integral of the QCD Lagrangian ({\it viz} the action).
The role of temperature in the Ising model is played by the couplings
in the QCD Lagrangian, the gauge coupling and the quark masses.
(Recall we are discussing zero temperature QCD simulations.
Finite temperature QCD simulations will not be discussed here.)

The Euclidean path integral still has one peculiarity not
present in the classical statistical system: the fermion fields are
anticommuting variables.  Formally integrating out the fermions, 
we obtain
\bea
\label{eq:EVpsi}
\langle J^\dagger(y)J(x) \rangle & = & \frac{1}{Z} 
\int [d\psi][d\overline{\psi}] [d U] ~J^\dagger(y)J(x)~ 
\exp\left(-\overline{\psi}\left(\gamma^\mu D_\mu + m\right)\psi 
~-~ {\cal S}_g\right)  \\
 & = & \frac{1}{Z} \int 
[d U] ~J^\dagger(y)J(x)~ \det Q[U]~ e^{-{\cal S}_g[U]}
\label{eq:EVdet}
\eea
where $Q=\gamma^\mu D_\mu + m$ is a matrix with spacetime, color, and
spin indices, and it depends on the glue field through the gauge-covariant
derivative $D$.  Finally, assuming we have been lucky or clever enough
to ensure that $\det Q > 0$, we can consider quantum observables like
(\ref{eq:EVdet}) to
be equivalent to statistical mechanical expectation values
like (\ref{eq:IsingEV}): the
sum over all possible field values, or configurations, 
weighted by a positive ``Boltzmann'' factor.

Now we are ready to use Monte Carlo methods to evaluate path 
integrals.  Since the Boltzmann weights are exponentials, most 
configurations give an exponentially small contribution to the integral.
Thus, the QCD path integral can be evaluated by generating an 
ensemble of ``important'' glue field configurations, ones
which maximize the Boltzmann weight.  One starts with
some initial configuration and uses an algorithm to create successive
configurations with a probability 
\beq
{\cal P}[U] ~=~ \det Q[U]~ e^{-{\cal S}_g[U]} \,.
\label{eq:weight}
\eeq
Using a finite number of configurations, $N$, gives rise to 
statistical uncertainty which decreases like $\sqrt{N}$.

The main obstacle lattice QCD calculations face is the difficulty in
computing $\det Q[U]$ for realistic quark masses.  The fastest algorithms
include the determinant by introducing wrong-statistics fermions $\phi$, 
such that 
\beq
{\cal P}[U] ~=~ \exp\left(-{\cal S}_g[U] ~-~ 
\phi^* \;Q^{-1}[U] \;\phi\right) \,.
\eeq
The cost of numerically inverting $Q$ increases (nominally)
like the ratio of the largest eigenvalue to the smallest, and
the smallest eigenvalue vanishes as the quark mass is taken to 0.
This, and other complications, make simulations with realistically
light quarks prohibitively expensive.

The valence, or quenched, approximation saves computer time by mutilating
the theory: the determinant is simply neglected in (\ref{eq:EVdet}) 
and (\ref{eq:weight}), and configurations are
generated with a weight determined solely from the gluon action.  
Consequently, effects of virtual quark loops like those
shown in Fig.~\ref{fig:corr} are omitted.  As a 
phenomenological model, quenched lattice QCD is not a bad one up
to the 10--20\% level (look ahead at Fig.~\ref{fig:ratio}).  
However, disagreement between calculation and 
experiment at this level leads to ambiguities
which cannot be removed within the quenched approximation.
In order to obtain the accuracy necessary to be relevant for
flavor physics, lattice simulations have to include sea quark effects.

\subsection{Improved Staggered Quarks}
\label{sec:impKS}

The improved staggered fermion discretization is
most expedient method to include light sea quark effects 
in present lattice QCD simulations.  Below a short description
of the main ideas behind staggering and improvement is given, followed
by an important caveat and my opinion of its relevance for phenomenology.

If we discretize the fermion Lagrangian by simply replacing
the derivative operator by a finite difference operator, we find
that the free massless fermion propagator,
\beq
G(p) ~=~ \frac{1}{i a \sum_\mu \gamma^\mu \,\sin\left(p_\mu a\right)} \, ,
\eeq
has poles not only at $p=0$, but also when any component $p_\mu = \pi/a$.
($a$ denotes the lattice spacing.)
The simplest solution, due to Wilson, is to add a term which
gives the extra 15 states, the ``doublers,'' a mass.  This term, however,
necessarily breaks chiral symmetry causing problems such as additive
mass renormalizations which make numerical simulation at small masses
intractable.  

\begin{figure}[t]
\begin{center}
\includegraphics[width=6cm]{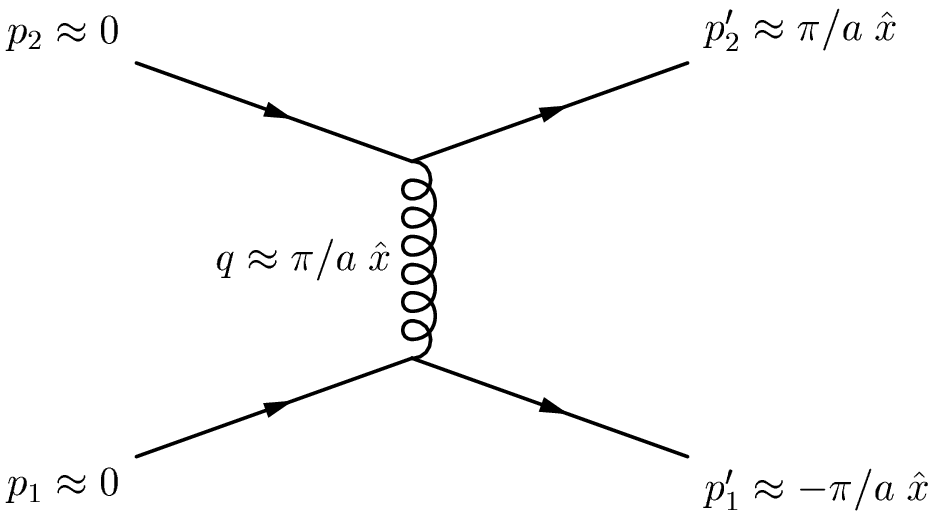}
\hspace{1cm}
\includegraphics[width=6cm]{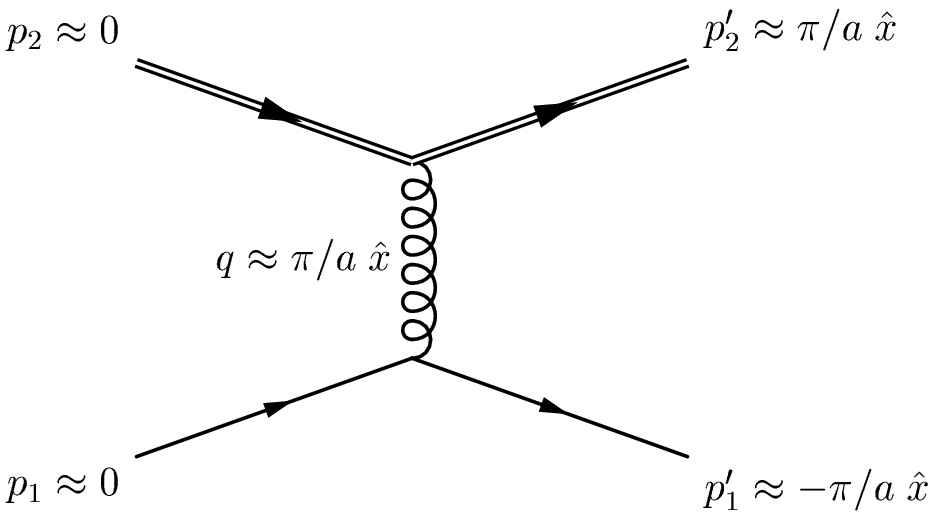}
\end{center}
\caption{\it \label{fig:qscatter} Quark-quark scattering by a gluon
with a large momentum component.}
\end{figure}

Instead of solving the doubling problem, the staggered formulation
embraces the extra fermions.
By using a lattice symmetry the 16 species are reduced to 4
which are interpreted as artificial flavors, called tastes.  Quarks of 
different tastes can interact by exchanging a gluon which has at least
one momentum component close to $\pi/a$ (see left diagram of
Fig.~\ref{fig:qscatter}).  The effect of this is that a low energy,
light meson propagator contains important contributions not just
from valence quarks with small momentum, but also from valence
quarks with large momentum components in opposite directions.
This mixing breaks the taste symmetry of the free fermion action and,
left untreated, leads to large discretization errors.  Based
on quenched studies it appears that without improvement, one 
would have to simulate
on lattices with spacings less that 0.06~fm 
to have control over discretization errors \cite{Bernard:1999xx}; 
if one also
requires the lattice volume to be bigger than $(2~\mathrm{fm})^3$,
simulations with light staggered sea quarks would not be feasible today.
Happily the leading taste-changing interactions
can be suppressed by modifying the lattice action 
\cite{Blum:1996uf,Lepage:1998vj}.  This improvement
allows one to obtain accurate results with attainable lattice
spacings and sizes, {\it e.g.}\ 
the coarse set of MILC collaboration lattices which have $a=0.13$ fm
and $V = (2.5~\mathrm{fm})^3$.

The news is even better for studies of heavy-light mesons.  In these
simulations, one uses a non-staggered discretization for the
$b$ or $c$ quark: either a nonrelativistic action or a Wilson-like
discretization.  Since the heavy quark formulation {\it does} 
get rid of the doublers, the heavy quark in Fig.~\ref{fig:qscatter}
(right) becomes very energetic when it absorbs or emits a hard 
gluon ($p_\mu \approx \pi/a$).  Such high energy effects give negligible
contributions to heavy-light meson propagators
\cite{Wingate:2002fh}.

\begin{figure}[t]
\begin{center}
\includegraphics[height=9cm,width=12cm]{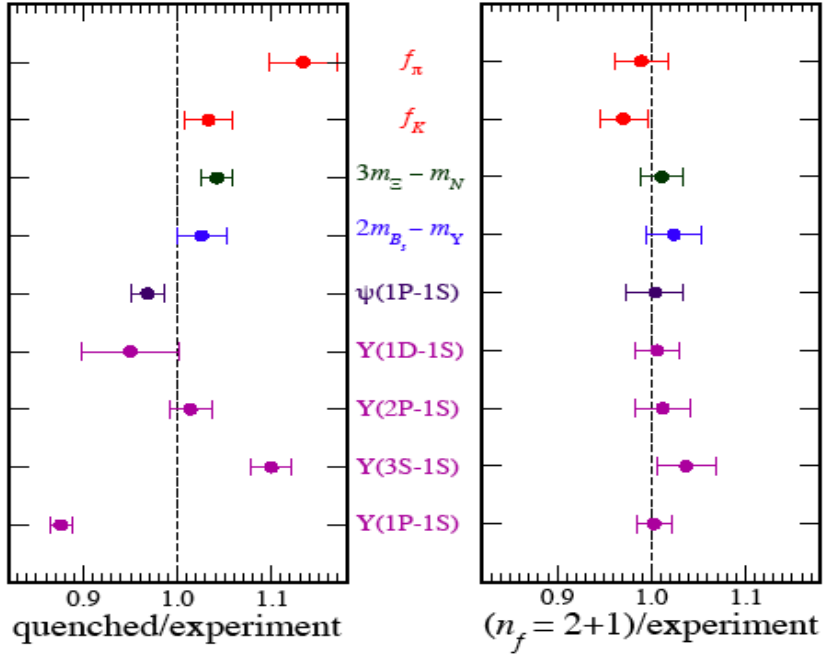}
\caption{\label{fig:ratio}\it Comparison
of quenched and unquenched results to experiment \cite{Davies:2003ik}.}
\end{center}
\end{figure}

Results using the improved staggered action for light sea quarks look
very promising \cite{Davies:2003ik}.  Figure~\ref{fig:ratio} shows
a series of lattice calculations divided by the corresponding 
observed physical values.
The quantities in the plot are some of the simplest ones to 
compute cleanly and correctly in lattice QCD.
\footnote{``Correctness'' here means that
we have a valid expectation for the simulation to agree with experiment.
The $\rho$ mass is a counterexample: it can be computed very cleanly
in lattice simulations; however, we should have no expectation
of obtaining $m_\rho = 770$ MeV until the simulated $\rho$ is above
the simulated $\pi\pi$ threshold.}
The contrast between quenched results (left) and unquenched results
(right) is striking.  

Alas, there is a fly in the ointment.  Even while we can tame the
taste-changing interactions, we cannot reduce the number of tastes
below 4.  Using a weight given by
\beq
\det Q_\mathrm{stag} \, e^{-S_g}
\eeq
produces an ensemble of configurations which include the effects
of 4 degenerate tastes of sea quarks.  In order to simulate a theory
with 2 light and 1 strange flavor of sea quark using staggered fermions
(improved or not), one takes a square root and a fourth root of the 
determinant
\beq
\left( \det Q_\mathrm{stag}^{m=m_{ud}} \right)^{1/2}
\, \left( \det Q_\mathrm{stag}^{m=m_s} \right)^{1/4} \, e^{-S_g} \, .
\label{eq:staggeredW}
\eeq
The problem is that the roots of the determinants cannot be
exponentiated to give a local fermion action.
The open question remains, does there exist a matrix $Q_\mathrm{local}$
which is QCD in the continuum limit ({\it i.e.}\ 
$\lim_{a\to 0} \overline{\psi}
Q_\mathrm{local}\psi = \overline{\psi}(\Dslash + m)\psi$)
such that
\beq
\left( \det Q_\mathrm{stag} \right)^{1/4} ~=~ \det Q_\mathrm{\,local} \, ?
\eeq
If so, then $2+1$ flavor staggered fermion simulations are {\it ab initio}
QCD calculations; if not, then the simulations merely model QCD.

In my opinion, this open question is an important one to study,
but it should not temper one's excitement over the success of
Fig.~\ref{fig:ratio}.  
We can leave behind the quenched approximation, 
a theory which we know is not QCD and disagrees with experiment, 
in exchange for light quark simulations which agree with experiment
but may or may not be formally QCD in the continuum limit.  
Enthusiasm for this approach is shared by
those outside the field (see e.g.~\cite{popular}).

Eventually nonstaggered lattice actions will replace staggered ones.
The beauty of lattice QCD is that, in the continuum limit,
it is formally QCD.  Presently the fourth-root trick spoils this beauty.
On the other hand, phenomenology dictates that results
be obtained with physical values for the quark masses, so
extrapolations are necessary.  Chiral perturbation theory tells
us how to extrapolate, but its domain of convergence is limited to
small quark masses.  With current and near-future resources, 
nonstaggered simulations barely overlap with the chiral regime.  
Consequently, the empirical extrapolations that are performed
using data at and beyond the border of the chiral regime 
are under no better theoretic control than staggered simulations 
which use the fourth-root trick.

\subsection{Miscellany}

Several important parts of lattice calculations
for flavor physics could not be addressed here.  
Nevertheless the interested reader should be
aware of work in these directions and may wish to consult the following
review articles.  (1)~Heavy quarks on the lattice cannot be treated the
same as light quarks when $m_Q a \ge 1$.  There are several 
approaches for treating heavy quarks on the lattice, and Kronfeld
discussed these last year~\cite{Kronfeld:2003sd}.
(2)~Much effort recently has been invested in fermion discretizations
which preserve the full flavor symmetries of the continuum, namely
{\it overlap} and {\it domain wall} fermions.  
These methods require significantly
more computer resources, so they presently cannot explore as deeply into
the chiral regime as staggered fermions.  Nevertheless, the full flavor
symmetry simplifies many analyses and no fourth-root trick is required
to simulate with 3 light flavors ({\it e.g.}\ see 
Ref.~\cite{Neuberger:2001nb} and therein). 
(3)~People are exploring {\it twisted mass} fermions, which
combine good features of both Wilson and staggered quarks and require no
fourth-root trick.  Frezzotti gave a review this 
year \cite{Frezzotti:Lat2004}.

\section{Some Recent Results}
\label{sec:results}

Many of the preliminary results reported below were presented
at {\it Lattice 2004}.  Since several months pass between the
conference and submission of proceedings, it is not uncommon in
our field to update results in the interim.  
For ``official'' Summer 2004 numbers, readers should consult
the authors' write-ups as they appear or Ref.~\cite{Wingate:Lat2004}.

\subsection{Heavy-light Decay Constants and $B^0 - \overline{B^0}$ Mixing}

\begin{figure}[t]
\ifx\pdftexversion\undefined
  {}
\else
  \vspace{-6cm}
\fi
\begin{center}
\includegraphics[width=14cm]{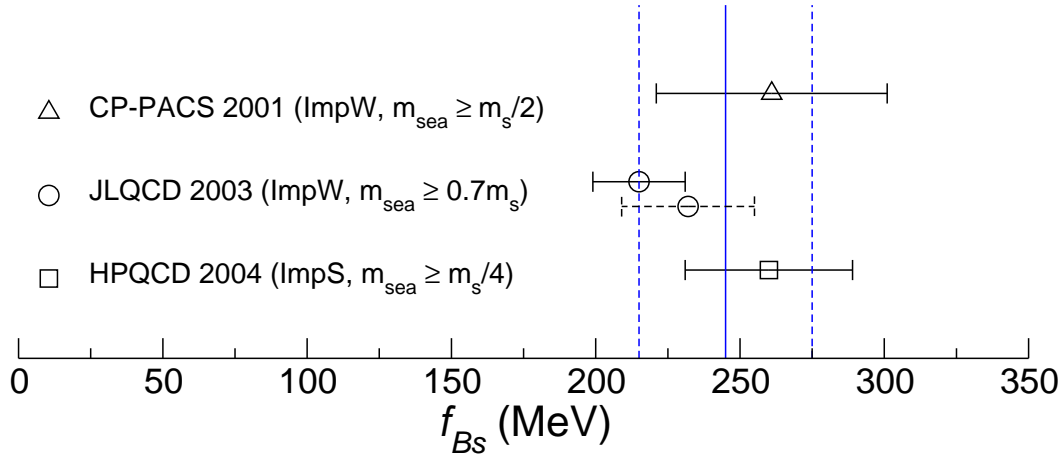}
\caption{\label{fig:fBs}\it Summary of unquenched $f_{B_s}$ calculations
\cite{AliKhan:2001jg,Aoki:2003xb,Wingate:2003gm}.
I have symmetrized the CP-PACS central value and averaged to obtain
the solid vertical line, 245 MeV.  
The dashed vertical lines denote 30 MeV systematic uncertainties.
The dashed point includes my estimate for the JLQCD result taking
into account the lattice spacing ambiguity observed by CP-PACS (see text).
}
\end{center}
\end{figure}

The $B_s$ and $D_s$ decay constants are relatively straightforward
to compute in lattice QCD since the strange quark mass can be tuned
to its physical value.  Figure~\ref{fig:fBs} shows 3 results 
 which compute $f_{B_s}$ on unquenched 
lattices \cite{AliKhan:2001jg,Aoki:2003xb,Wingate:2003gm}; 
the first 2 use improved Wilson fermions as light as $m_s/2$, 
and the last uses improved staggered fermions as light as $m_s/4$.  
Ref.~\cite{AliKhan:2001jg} observed a 10-20\% lattice scale 
ambiguity, which they include in their error estimate.  
Ref.~\cite{Aoki:2003xb} did not compute the $\Upsilon$ splittings,
but given the similarity in the sea quark mass range simulated,
I suspect that they will see a similar effect; I indicate my 
prejudice with the dashed point in Fig.~\ref{fig:fBs}.  See
Ref.~\cite{Wingate:Lat2004} for a more detailed argument.
A straight average of the 3 published results (shifting the
CP-PACS central value so that the error bars are symmetric) gives 
$f_{B_s} = 245$ MeV.  Since all 3 calculations have similar dominant
systematic errors (due to truncating perturbative expansions) averaging
does not reduce this uncertainty: 30 MeV is the typical estimate
of these truncations.

Refs.~\cite{AliKhan:2001jg,Aoki:2003xb} give unquenched results for
$f_B$ (and the ratio $f_{B_s}/f_B$ which gives a more restrictive
contraint on $|V_{td}|$).  However, in my biased opinion,
unquenched simulations with light quark masses below $m_s/2$ are
necessary to have a trustworthy overlap with chiral perturbation
theory.  Work using improved staggered fermions is underway, and
progress is reported in Ref.~\cite{Gray:Lat2004}.

\begin{figure}[t]
\begin{center}
\includegraphics[width=6.5cm]{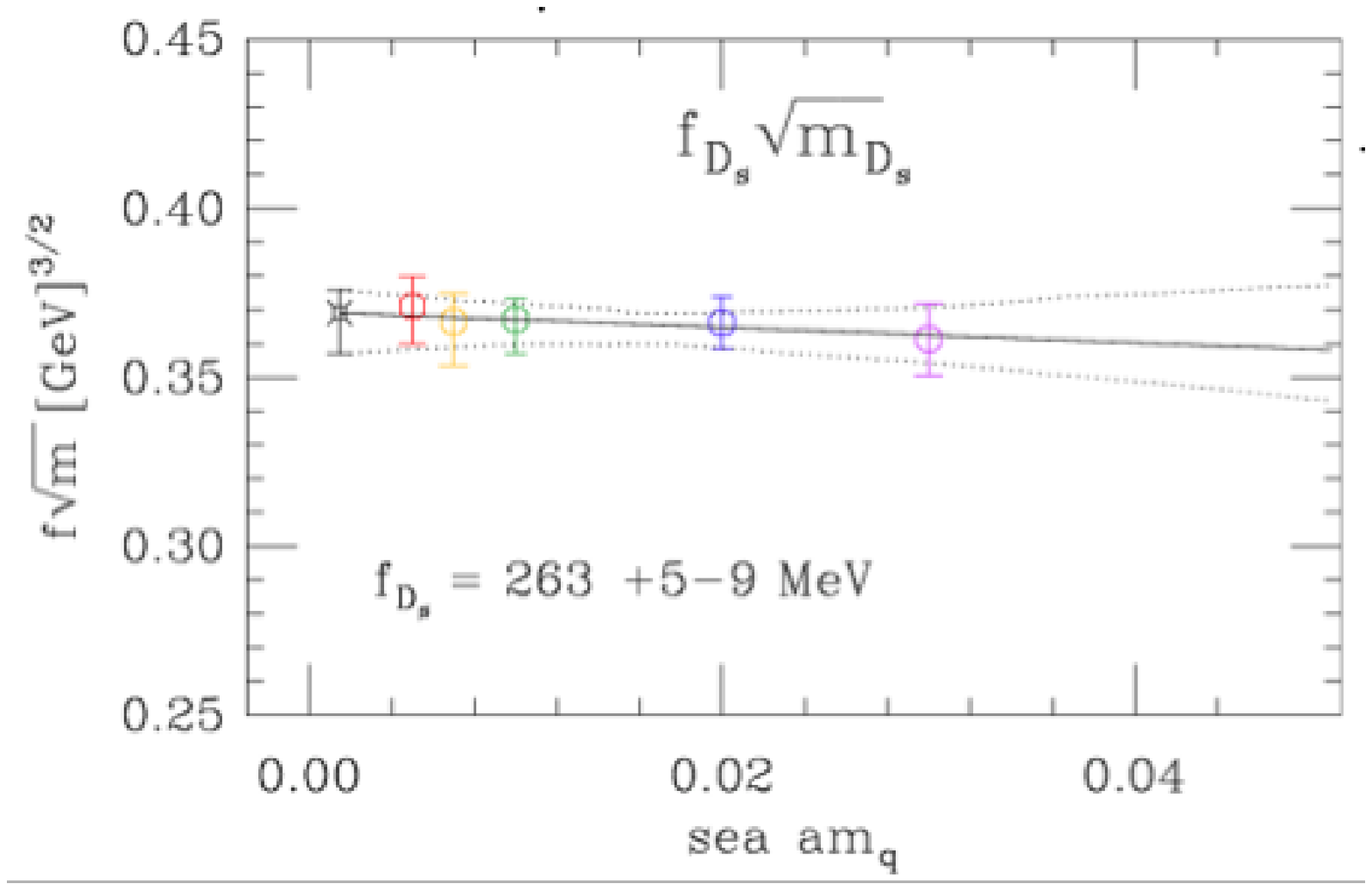}
\hspace{1cm}
\includegraphics[width=6.5cm]{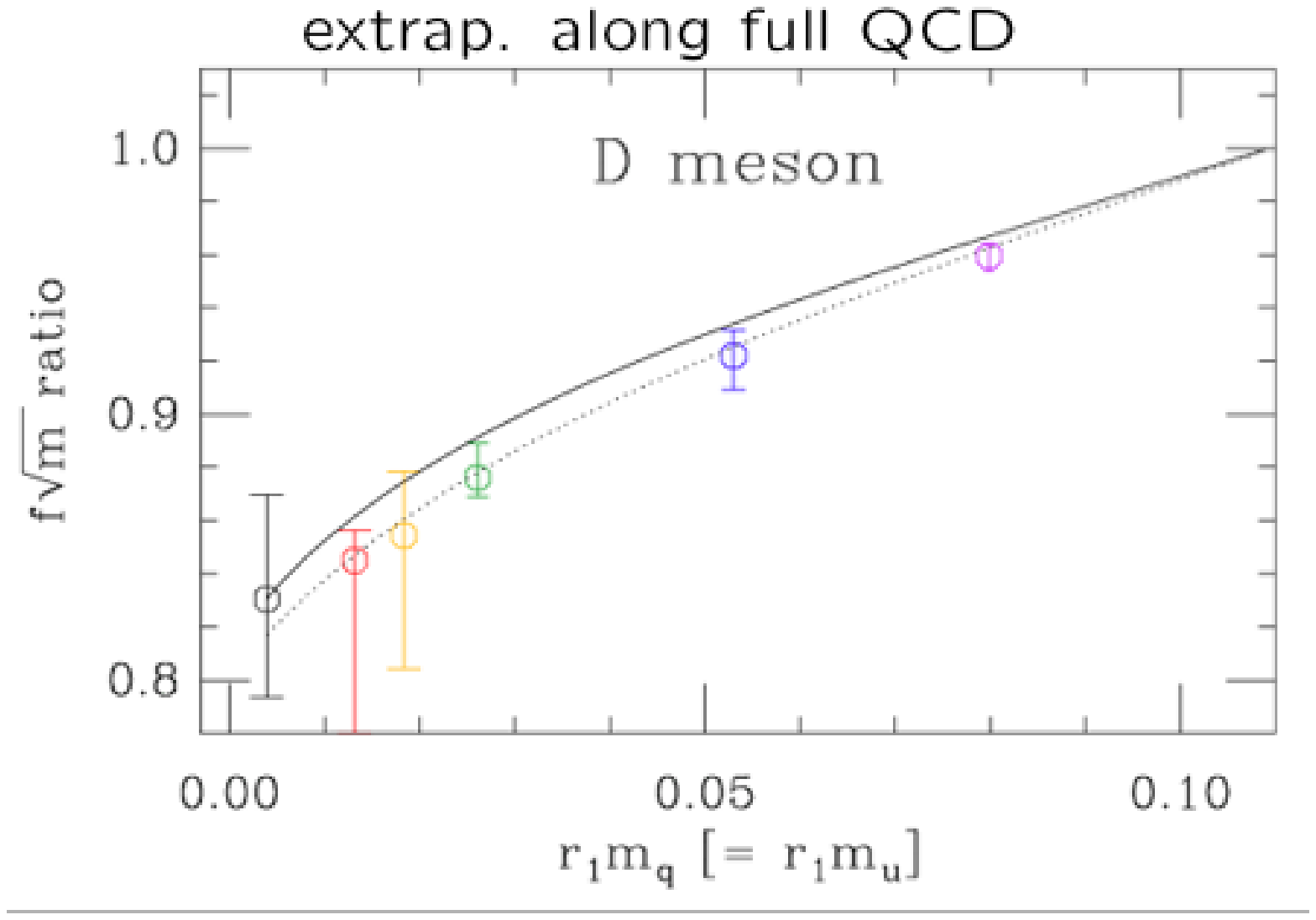}
\end{center}
\caption{\label{fig:fD} \it Preliminary Fermilab/MILC calculation
of $f_{D_s}$ (left) and $f_D/f_{D_s}$ (right) \cite{Simone:Lat2004}.
The physical strange quark mass corresponds to $am_s = 0.04$ and
$r_1 m_s = 0.11$.  On the right, the solid line is a fit to
chiral perturbation theory which includes finite lattice spacing
effects.}
\end{figure}

Figure~\ref{fig:fD} shows preliminary results for the $D_s$ and
$D$ decay constants computed using Fermilab heavy quarks and improved
staggered light quarks on the MILC configurations \cite{Simone:Lat2004}.
Since they have a large set of ``partially quenched'' data 
(where the valence quark mass is allowed to be different than the
sea quark mass), they can do sophisticated global fits to partially
quenched chiral perturbation theory, even including leading
taste-breaking effects \cite{Aubin:Lat2004}.
I will delay quoting results until Ref.~\cite{Simone:Lat2004} appears,
but the dominant uncertainty in $f_{D_s}$ and $f_D$ has been 
estimated to be 10\% from the heavy quark matching to QCD.  This
uncertainty largely cancels in the ratio, so they anticipate
the dominant uncertainty for $f_{D_s}/f_D$ to be 5\% 
coming from statistics and the chiral fits.

The $\Delta B = 2$ hadronic matrix elements relevant for 
the $B^0-\overline{B^0}$ mass and lifetime differences can be
directly calculated, although the numerics are more difficult
than for the decay constant.  JLQCD has calculated these with
2 flavors of dynamical improved Wilson fermions in
\cite{Aoki:2003xb,Yamada:2001xp}.
Calculations using improved staggered fermions are underway
\cite{Gray:Lat2004}.

\subsection{Semileptonic $B$, $D$, and $K$ Decays}

\begin{figure}[t]
\begin{center}
\includegraphics[width=7.2cm]{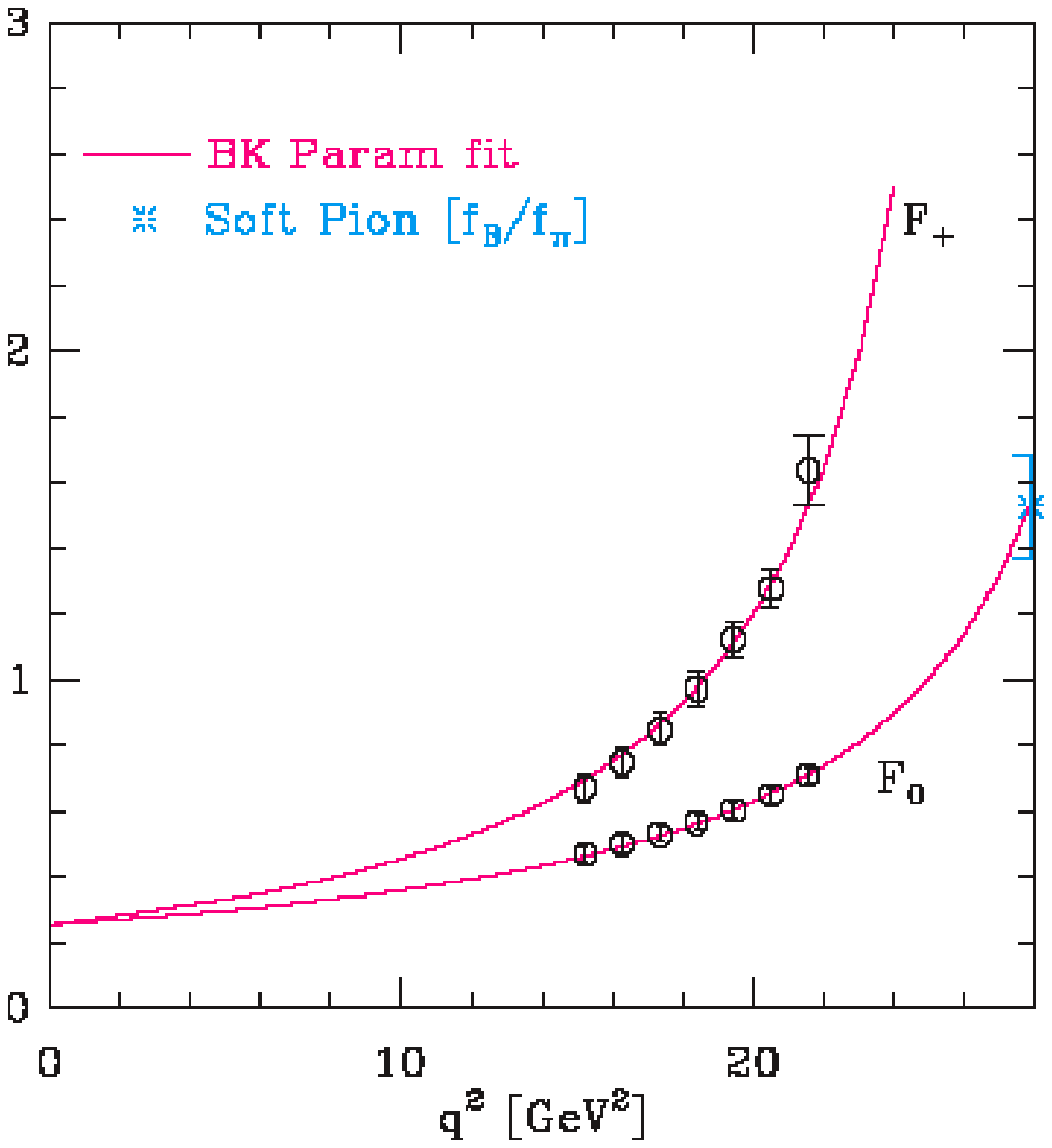}
\hspace{0.3cm}
\includegraphics[height=7.3cm,width=7cm]{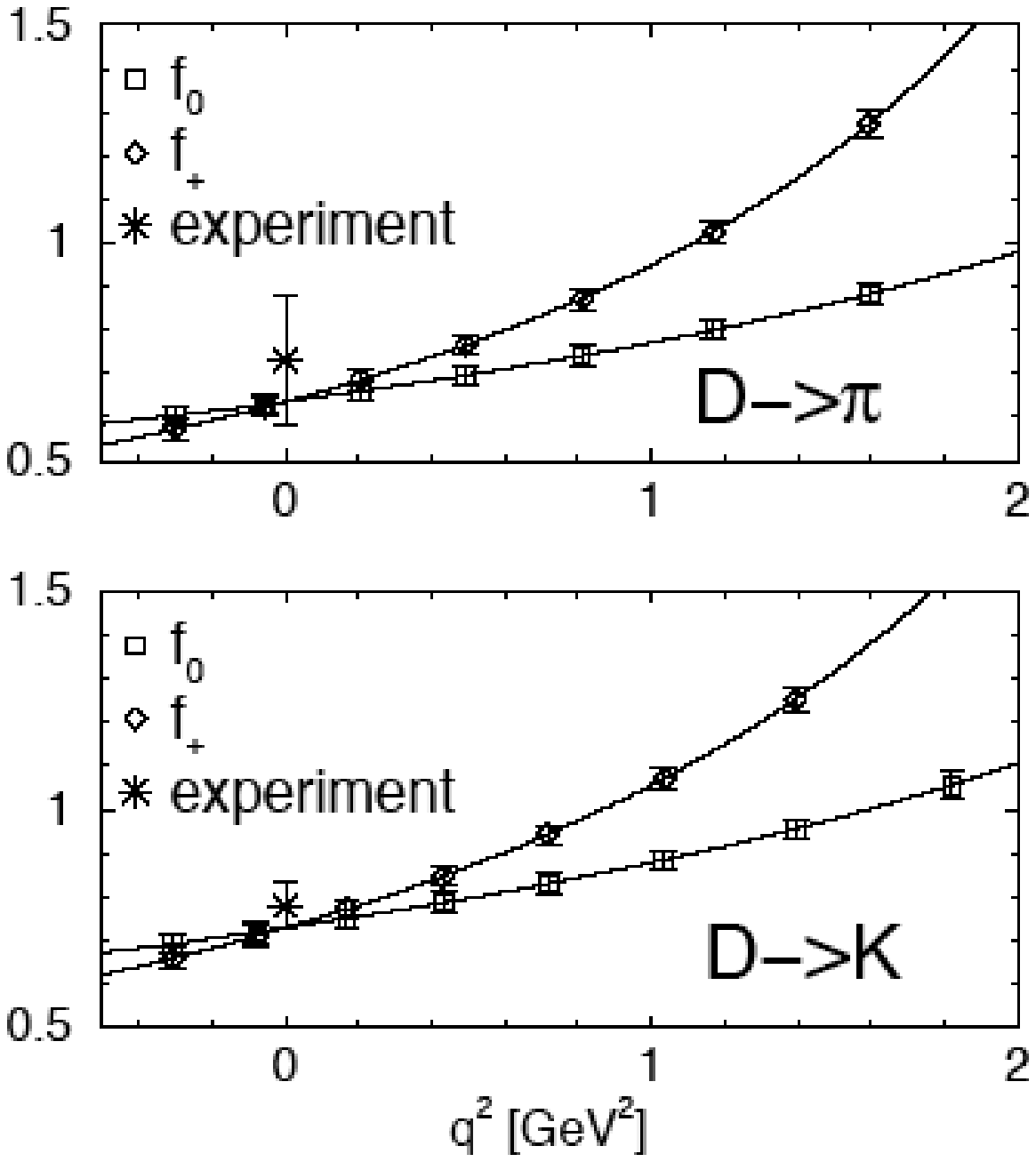}
\caption{\label{fig:FF}\it $B\to \pi\ell\nu$ form factors (left)
\cite{Shigemitsu:2004ft},
 and $D\to\pi\ell\nu$ and $D\to K\ell\nu$ form factors (right) 
\cite{Aubin:2004ej}.}
\end{center}
\end{figure}

Calculations of the semileptonic form factors parameterizing
$B$ and $D$ semileptonic decays have recently been carried out
on the unquenched MILC configurations 
\cite{Shigemitsu:2004ft,Aubin:2004ej}.  The form factors
$f_+$ and $f_0$ are shown as functions of $q^2$, the momentum 
carried away by the lepton pair, in Fig.~\ref{fig:FF}.  The dominant
uncertainties in these calculations come from  the truncation of
the heavy quark effective action.

The data are fit well by the Be\'cirevi\'c-Kaidalov ansatz
\cite{Becirevic:1999kt}, which is used to extrapolate (for $B$)
or interpolate (for $D$) to $q^2 = 0$.  Furthermore one can
integrate $f_+$ over $0\le q^2 \le q^2_\mathrm{max}$ and combine the 
result with experimental branching ratio and lifetime
to determine the corresponding
CKM matrix element.  In the case of $B\to\pi\ell\nu$, one cannot
presently simulate at small $q^2$ without inducing large discretization
effects.  Restricting both the form factor integration and the
branching fractions to decays with $q^2 \ge 16$ GeV${}^2$ 
\cite{Athar:2003yg} reduces
the theoretical error in $|V_{ub}|$ but increases the experimental 
error~\cite{Shigemitsu:2004ft}.

\begin{figure}[t]
\begin{center}
\includegraphics[height=7.5cm,width=7.0cm]{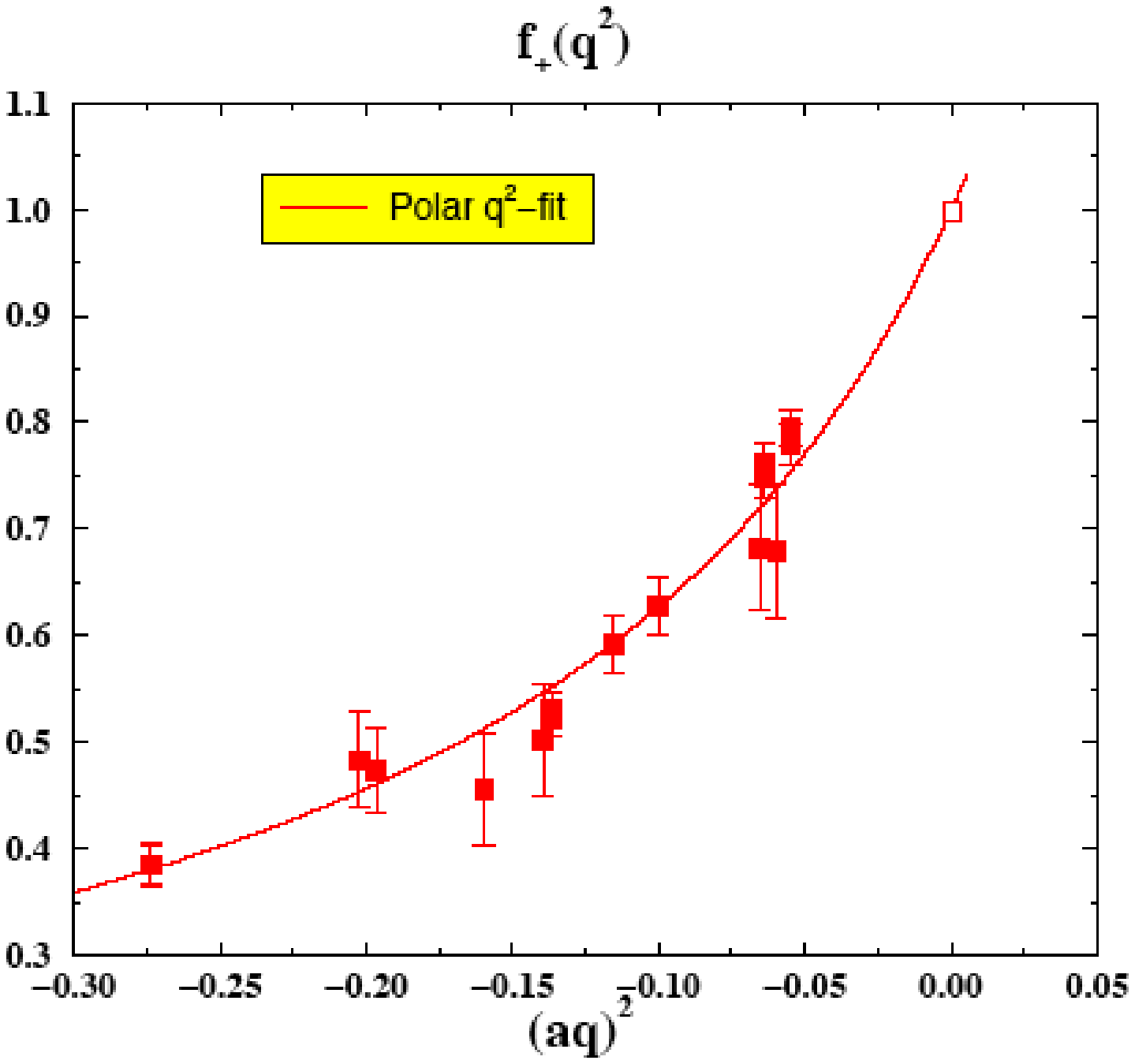}
\hspace{0.1cm}
\includegraphics[width=7.5cm]{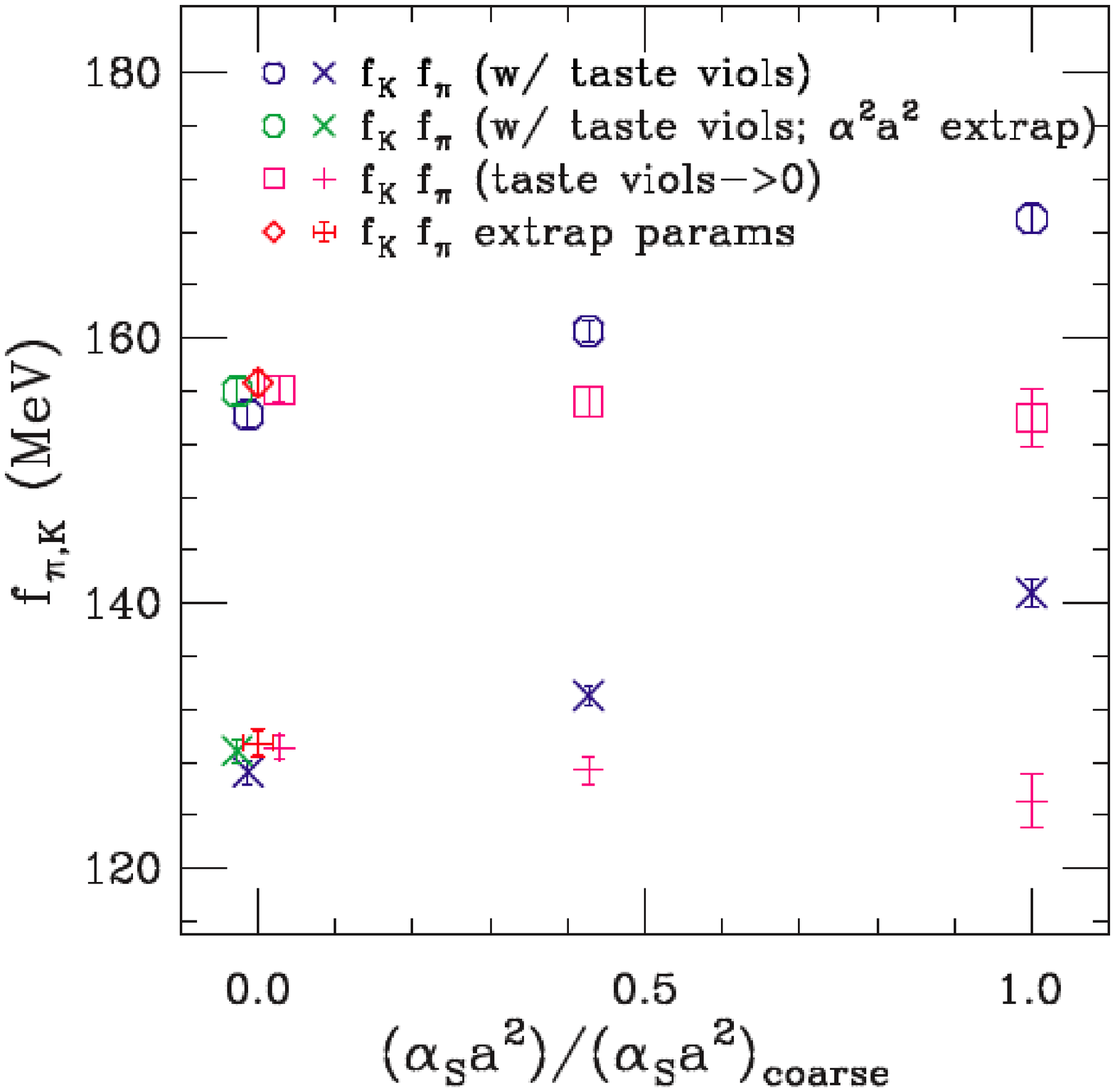}
\end{center}
\caption{\label{fig:Vus} \it $K$ semi-leptonic form factor
$f_+$ vs.\ momentum transfer \cite{Becirevic:2004ya} and 
$K$ decay constant (upper points) vs.\ lattice spacing
\cite{Aubin:2004fs}.}
\end{figure}

Unprecidented precision is being obtained in lattice calculations
of $K$ decays.  Figure~\ref{fig:Vus} shows the $K\to\pi\ell\nu$
form factor $f_+(q^2)$ in the quenched approximation (left) 
\cite{Becirevic:2004ya}
and the unquenched $K$ (top right) and $\pi$ (bottom right) 
decay constants \cite{Aubin:2004fs}.  The calculation of $f_+$ is
notable for several technical innovations which allow a signal
to be extracted.  The calculation of $f_K/f_\pi$ is now precise
enough lead to a value for $|V_{us}|$ competetive with the
semileptonic determination 
\cite{Marciano:2004uf,Aubin:2004fs}.

\subsection{Quarkonia and $B_c$ Masses}

We have already seen in Fig.~\ref{fig:ratio} the improvement
in the bottomonium spectrum when light sea quark effects are included
\cite{Davies:2003ik}.  Preliminary charmonium results were presented
last year \cite{diPierro:2003bu}.  Work is underway to finalize these
results.

\begin{figure}[t]
\begin{center}
\includegraphics[height=4.5cm,width=7.0cm]{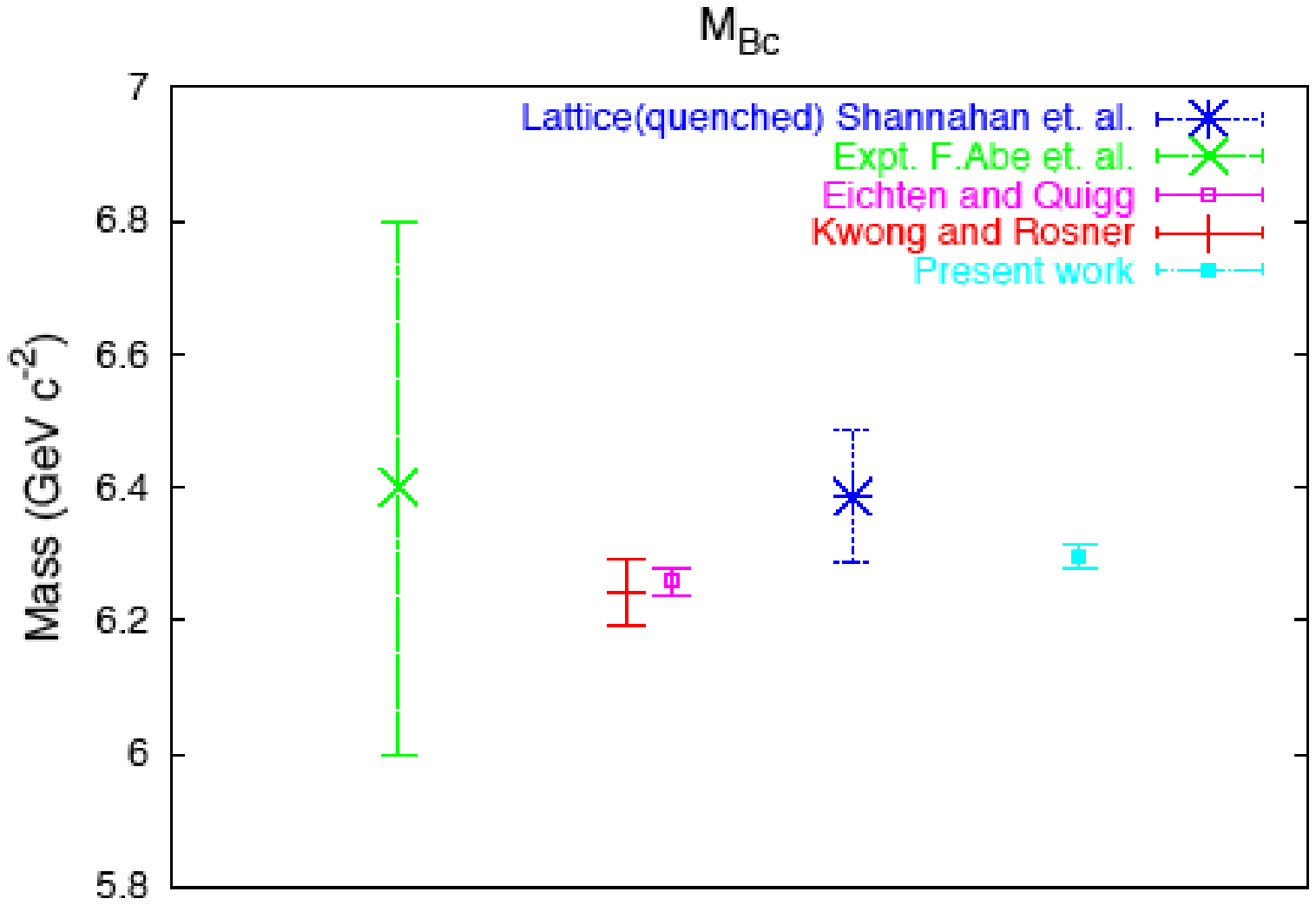}
\hspace{0.1cm}
\includegraphics[height=4.6cm,width=7.0cm]{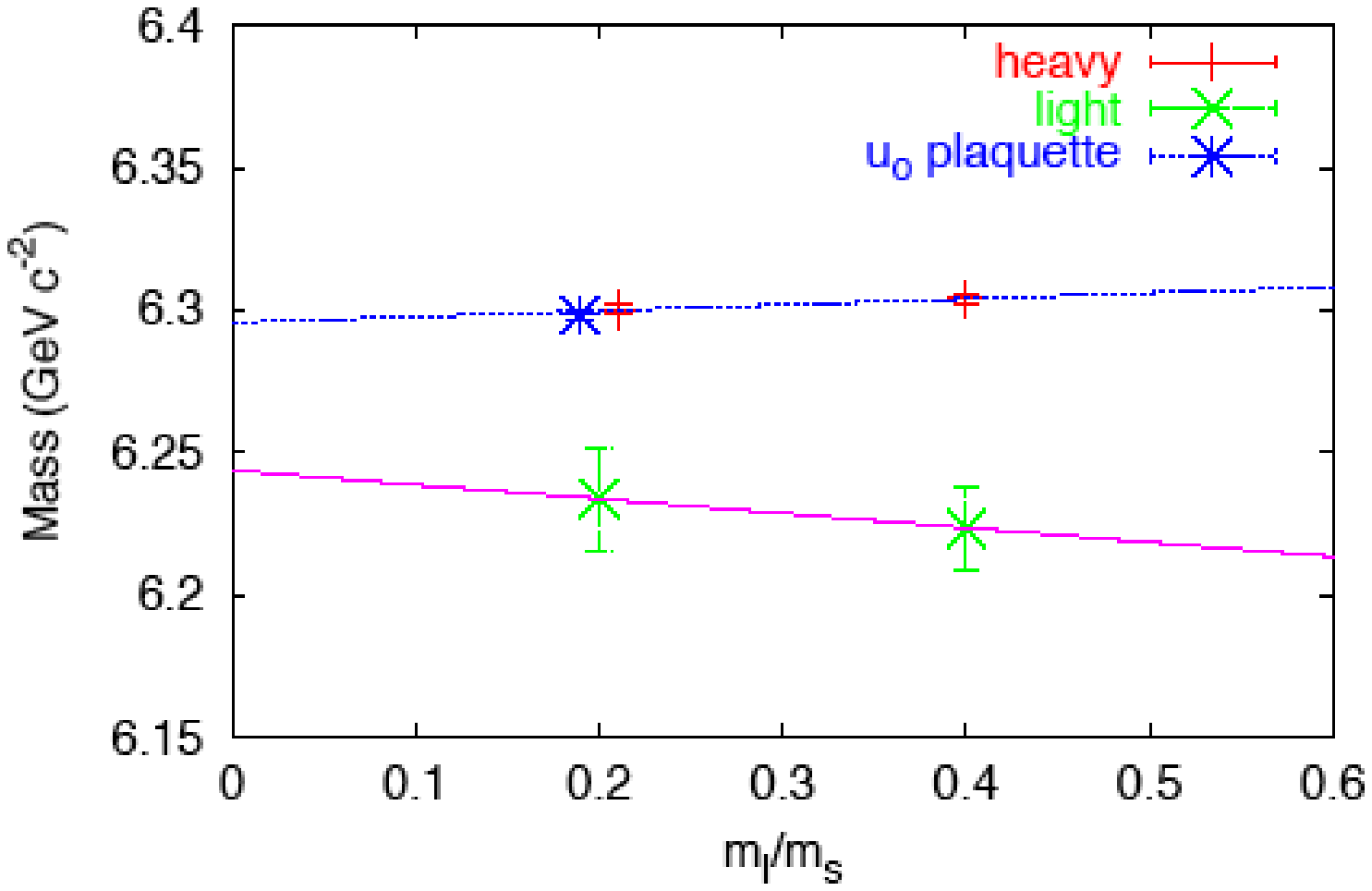}
\end{center}
\caption{\label{fig:Bc} \it Left: $B_c$ mass from experiment (far left
point), models (2nd and 3rd points), and lattice (4th point quenched,
5th point unquenched) (see \cite{Allison:Lat2004} and Refs.\ within).
Right: Comparison of $m_{B_c}$ computed using two mass differences (see
\cite{Allison:Lat2004} and the text); results agree within errors.}
\end{figure}

The $B_c$ meson mass has been computed using the unquenched MILC
configurations \cite{Allison:Lat2004}.  Fig.~\ref{fig:Bc} (left) shows
the result as the rightmost point and comes from computing 
$m_{B_c} - (m_{J/\psi} + m_\Upsilon)/2$ on the lattice and using
the experimental quarkonium masses.  Fig.~\ref{fig:Bc} (right, upper points) 
shows this mass difference vs.\ sea quark mass as well as a check
using the difference $m_{B_c} - (m_D + m_B)$ (lower points);
only statistical errors are plotted.  The systematic uncertainties
are also larger using the $B$ and $D$ --
about 50 MeV compared to 10 MeV for the mass difference from
$J/\psi$ and $\Upsilon$ -- so while the 2 results agree, the more
precise one is taken for the preliminary result.
Given the precision of the
lattice result compared to experiment, this will be an interesting
prediction to compare to Tevatron Run II data.

\section{Conclusions}
\label{sec:summary}

In my opinion the next few years will be exciting ones for lattice
QCD.  Having shed the quenched approximation, lattice
calculations can reach enough precision and accuracy to help 
constrain the CKM matrix elements.  In order to do so, hard work
is still required to reduce uncertainties from extrapolations
in quark mass, finite spacing and size effects, and operator matchings.
In the absense of a theoretical proof, the open question behind the 
staggered fourth-root trick will hang over our heads.  Even so,
the accumulation of empirical agreement between staggered simulations,
experiment, and eventually non-staggered simulations, is enough to
have give important contributions to phenomenology.  Finally, 
there are still
classes of problems which are difficult to address with current
techniques -- decays with multiple final-state hadrons, in particular --
which will receive much more attention if the simpler problems
can ever be said to be done.

\section*{Acknowledgments}

I thank I.~Allison, F.~Mescia, M.~Okamoto, J.~Simone, and my collaborators
 for sending me some of the material
presented here. I acknowledge helpful conversations
with C.~Davies, S.~Hashimoto, S.~Sharpe, and J.~Shigemitsu.


\end{document}